# Direct Fractional Auction*

By

Dmitriy Taubman
dmitriy.taubman@gmail.com

Boris Gleyzer
boris@deltaonefutures.com

Ke Yang
kyang@hartford.edu

Farhad Rassekh**
rassekh@hartford.edu
Department of Finance, Analytics, Risk, and Economics
Barney School of Business
University of Hartford
West Hartford, CT 06117

November 2024

**Abstract**

This paper designs a market algorithm for fractional ownership of an indivisible asset. It provides an efficient market mechanism, named Direct Fractional Auction (DFA) that offers valuable assets to both small and large investors who can become partial owners of such assets. Additionally, it introduces procedures and algorithms with DFA to determine the optimal winning combinations of unaffiliated bidders. DFA, on the one hand, transfers the partial ownership to the winners and, on the other, redirects the proceeds from the auctions to the sellers. We show that the DFA algorithm works more efficiently than the Greedy algorithm in maximizing the seller's value. We also demonstrate the possibility of reducing the complexity of the problem using the "pruning" method for data pre-processing.

# Direct Fractional Auction

## 1. Introduction

Partial or fractional ownership of an asset through auction has become increasingly common in financial markets, and spurred exponential research in this area. Wilson (1979) is a seminal work that compares the sale prices resulting from a share auction with those from a unit auction and finds share auction can yield a significantly lower sale price. Meanwhile, a multi-object auction, often called "combinatorial" auction, can be considered a more general form of the share auction. Nisan (2020) suggests that an approach based on linear programming (LP) can be used to find the optimal solution for the combinatorial auction. Bogyrbayeva et al (2020) use an iterative combinatorial auction design for a new market for fractional ownership of Autonomous Vehicles (AVs), in which an AV is leased by a group of individuals. In the last decade, blockchain technology has gained increasing attention from the public, policymakers, and investors. This technology is widely used in the financial sector, management, and economics (Nobanee and Ellili, 2022). More specifically, a growing number of research papers examine the connection between Non-Fungible Tokens (NFTs) pricing and other asset pricing, such as real estate (Dowling, 2022a), cryptocurrencies (Dowling, 2022b), and major and financial assets (Umar et al., 2022).

Direct Fractional Auction (DFA) is a new blockchain-based auction-market-design mechanism that enables investors to bid directly for a portion of an asset and to co-own this asset with other unaffiliated investors. It is practically applicable to any asset: real estate, artifacts, paintings, horses, rare collectibles, and so on. One of the more convenient ways to accomplish such a task is through fractionalized NFTs (Non-Fungible Tokens) representing any such asset. Most NFT auctions are conducted on a "winner takes all" basis, where the single highest bid wins. In some cases, private investors form groups to buy a whole NFT on a negotiated basis, but this approach does not allow for a more efficient open market price discovery. At the other end of the spectrum, NFTs may be securitized, broken into many shares, and sold to investors through the Over The Counter (OTC) - type markets where thousands would potentially become small shareholders.   In contrast, Direct Fractional Auction provides a simple and efficient mechanism to allow shared ownership of NFTs and other goods. It creates a new type of marketplace that allows unaffiliated investors to bid directly for a portion of an asset and for its total ownership. Direct Fractional Auction rests on the idea that dividing NFT into a predetermined number of partial ownerships (Fungible Tokens) enables more small investors to become part owners of an asset that may otherwise be too expensive for each of the buyers while limiting the total number of potential owners, thus creating meaningful ownership of an asset.

While there are certain similarities between the DFA and Multi-Unit auctions (combinatorial auction) – and the selection problems of winners could be mathematically resolved – economically these two types of auctions are entirely different.  DFA auctions provide a path to a meaningful, continuous, fractional ownership of valuable assets with a possibility of long-term investment and availability of secondary markets. In addition, DFA auctions provide a chance for full ownership, if desired, by buying all issued FTs and converting them back into the full original NFT.   The DFA approach is fully compatible with the existing process of – well-known and widely used – fractionalizing NFTs.   Its capabilities include minting new NFTs if needed or working with previously issued NFTs, then converting NFTs into multiple Fungible Tokens (FTs) and auctioning the FTs through the DFA directly to the qualified auction participants.

Each NFT offered for DFA is divided into a predetermined number of Fungible Tokens.

When an NFT is fractionalized, it is first locked into a smart contract, which also defines the terms of ownership of FTs. The smart contract then splits the NFT token into multiple fractions, each representing partial ownership of the NFT. Investors will own a fraction of the NFT equal to the number of their tokens divided by the total number of fungible tokens produced when the NFT was locked in a contract. For an illustration, suppose we divide an NFT into K equal ownerships of fungible tokens (FTs). Auction participants can bid for any number of Fungible Tokens from 1 to K. Direct Fractional Auction bidding requires the specification of two parameters: (1) the number of FTs, and (2) the price the buyer agrees to pay for them. A bidder for w number of FTs, where $1 \leq w \leq K$, agrees to pay the price for only their total number — similar to the "All or None" order type in a regular exchange. Such an auction aims to obtain the highest possible price for the seller. The main challenge in determining the auction winners is to ensure that the total number of FTs sold does not exceed K while providing maximum benefit to the seller. The mathematical model required for such an auction is the well-known Knapsack Problem. We develop a unique mathematical model that provides a novel solution for this version of the Knapsack Problem, thus accomplishing the goal of the auction: to sell partial shares of the NFT to the new unaffiliated owners while maximizing the total selling price for the seller.

## 2. Advantages of the Direct Fractional Auctions (DFA)

DFAs of NFTs offer a variety of benefits compared to traditional trading in NFTs, some of them are:

- Small investors can own a share of an asset that would otherwise be prohibitively expensive. Existing markets for expensive NFTs often have low liquidity, few participants, and high transaction costs. Such NFTs sometimes are split into multiple fungible tokens (fractionalized) and traded on OTC exchange-type markets such as double auctions. This increases the number of small investors and improves liquidity. Such markets are like stock markets with a relatively large number of shareholders, which means any given small investor lacks meaningful ownership. DFA auctions will attract an increased number of participants by attracting smaller investors (bidders) while still limiting the ultimate number of co-owners.

- Diversification allows investors to reduce risks by investing in various NFTs. DFA enables individuals to invest in multiple NFTs and still be a meaningful, although partial, owner of an asset.

- More efficient price discovery. Valuation of rare goods – such as unique artworks that do not have an extensive price history – has proven to be difficult. In this case, a larger number of investors in the open market can determine a price better than an individual investor or group of affiliated investors.

- Direct Fractional Auction potentially maximizes sellers' revenue. It has more bidders under conditions similar to those of other types of NFT auctions. By allowing participants to bid for a portion of an NFT (or any other tangible asset) for sale and collective ownership, DFA lowers the participation threshold in the auction, essentially democratizing the auction market and attracting more bidders. Note that DFA also allows bidding for the whole asset by entering a price for the entire predetermined number of fungible tokens. Both groups – partial ownership bidders and whole ownership bidders – can participate in the auction, potentially resulting in a higher price for the seller.

3. Details of the Closed Fractional Auction -- Selection of Winners

At a closed (sealed bid) auction, with the help of an auctioneer, sellers put up an item, which potential buyers get acquainted with before making bids. The sellers set a reserve price $r \geq 0$, below which they do not agree to sell their items (the reserve price may be zero). The auction takes place in the time interval from the start, which is announced in advance, until the fulfillment of the conditions set by the auctioneer. Each bidder has the right to make an unlimited number of bids. All information about bids is delivered by participants in a closed form and becomes known only after the end of the auction. In the fractional auction, the bidders need to specify two numbers: First, the number of fungible tokens they bid for. Second, the price they are willing to pay for the share (number of tokens). The total number of FTs in the winning group must not exceed K, its maximum specified number. This is the basic condition for collective ownership of an asset. The criterion is to maximize the sum of the auction winners' prices, thus providing the sellers with the highest price for their assets.

Although the DFA auction is designed as a closed (sealed bid) auction, there are built-in capabilities to periodically display a group of anonymous bids with the highest combined price during the auction. The seller decides whether to make this information available to the participants both visually on the screen and through the API (application programming interface) for computerized bidders. This could improve the liquidity of the auction and create no opportunity for arbitrage. The current version of the DFA has been implemented on the Ethereum blockchain platform, which is also the basis for most NFTs projects. Further implementation will include additional platforms, such as Tezos, a decentralized blockchain proof-of-stake protocol.

## 4. Fractional Auction with a Limit on the Number of Fungible Tokens

Let us consider fractional auctions, where all bids are described by ratios of two integers of the form w/K, where $1 \leq w \leq K$ and K is the maximum specified number of FTs. The auctioneer decides how many equal parts underlying NFT are divided into (number) K. This number is communicated to all participants before the start of bid collection and is a guideline for the formation of bids. Since for a fixed auction the number K is constant, the share w/K of each bidder is determined by choosing the integer w. For example, if K = 12 and the bidder wants to have 25% of the shares, then his fungible tokens w = 3. Therefore, the shares are measured in whole numbers. The unit is always the minimum fraction, and in this example, the number 12 is the maximum number of tokens which would be the bid if a bidder wants to buy the entire NFT. The main constraint for determining the group of winners in this problem is that the total shares owned by all members of the group must not exceed K. Thus, in a fractional auction with a common denominator K, the goal of selecting winners is to find a group of bidders for which:

- The sum of all FTs owned by the group members does not exceed K.
- The combined amount they will pay is the maximum possible.

The auction generates the list of winners whose total value of the collected payments is transferred from the winners (K*) to the seller. Winners are issued certificates of ownership, which indicate the share of each w of the total number of the winning FTs (K*). If the result is that K*< K, a strict inequality, then the calculated total number of the FTs of the winners (K*) replaces the initial number K in the certificates that are issued to the winners. At the same time, the new share w/K* of each winner becomes greater than the share w/K that they bid for. This is a benefit for the auction participants as the market determines the organically true value of the NFT and the true share of Fungible Tokens.

Note that if there is sufficient liquidity, then the equality K* = K takes place. For example, this will happen if the number of bids with w = 1 is equal to or greater than K. The number K, which is the number of FTs that the underlying NFT is divided into, plays a vital role in the organization of the auction. Since each potential owner contributes a bid of no less than 1

FT, and the total number of FTs of future owners must not exceed K, then the number of future owners cannot exceed K.

## 5. Fractional Auctions Problem with the Condition of Uniqueness of the Solution

The mathematical model of choosing the winners of a fractional auction with a limit on the number of shares is similar to the well-known integer 0_1 knapsack problem. The number K sets the size of the Knapsack, and significantly affects the complexity of algorithms for choosing winners, which are based on recursion and algorithms such as dynamic programming. Ensuring clarity when choosing a group of winners, especially for closed auctions, there may be cases where several groups of bidders collectively offer the same price, and then additional rules are essential for the selection of the winning group. For this purpose, it is required to use another parameter, t - the time the bid enters the system. Thus, each group G, which includes several bids $G = \{b(1), b(2)… b(s)\}$, where each bid is associated with a particular time when it was placed, i.e.: bid b(1) was placed at the time t1, bid b(2) at the time t2 and so on. We assume that two bids cannot enter the system at the same time. The process that determines the winning group of unaffiliated bidders first selects a group of bids that combine to give the highest price for the seller. If there are two or more aggregate bids that are equal in price, the priority is given to the earliest one. This is determined by comparing times of the last (latest) bids in the competing groups. If the last bids are the same, then the comparison moves back to a bid before last, and so on. (Appendix 1 provides an example).

Thus, the bid groups are ordered in lexicographic order by t, like words in a dictionary, where times t(i) replace letters, and alphabetical order is replaced by the natural order on the t-axis. This order is strict and therefore has the smallest element that uniquely defines the group of winners. To obtain a unique solution, we must add an additional condition to the formulation of the knapsack problem: ensuring the earliest arrival time for the group associated with the maximum cost.

## 6. Defining the mathematical problem of selecting the winning group

**Input data:**

Suppose there are n bids B = {$b_i$ = ($w_i$, $p_i$)}, i = 1……n}, where

$w_i$ is the number of tokens entered for $b_i$. $w_i$ is a whole number;

$p_i$ is the price for $w_i$ tokens entered for $b_i$. $p_i$ is a positive number;

$t_i$ is the time stamp of the bid i when it is accepted be the system;

K is the specified number of fungible tokens.

Domains: $p_i \in \mathbf{R}, \mathbf{R} > 0$, $w_i \in \mathbf{N}^* = \{1,2,3, 4...\}$.

**Problem**:

Find the subset of bids M, M ⊂ B, where B is all given bids, such that:

$$\sum_{i \in M} W_i \leq K$$

and

$$F = \sum_{i \in M} P_i \Rightarrow \text{MAX}.$$

In the case of multiple solutions, it is also required that the unique optimum solution M* be the smallest in the lexicographic order in t. With a very large number of participants, we apply data pre-processing which we call "pruning" by using a method described in Appendix 2.

For a possible solution of the problem formulated above, we first check the traditional Greedy algorithm using an approach that allows us to calculate the price/weight ratio for each item and then sorting the items based on this ratio. Let us consider a simple example, using an algorithmic approach for the solution of the problem:

K (number of all FTs) = 2, and three bids received: b1 = (2, 4), b2 = (1, 3), b3 = (1, 0.5).

The solution obtained through Greedy algorithm would be: {b2, b3}; however, the true solution for M*= max must be {b1}. Thus, in this example the Greedy algorithm approach does not produce the correct result for the problem since the seller does not receive the maximum amount. In general, Greedy will not always arrive at a correct solution when the problem is defined as a 0_1 Knapsack problem, and it will not be able to maximize the seller price in every case as shown above. The proposed DFA algorithm resolves this issue.

We sort the input arrays b(i) and w(i) according to the time of receipt of bids in the system. This means that their order, index i, satisfies the condition: t(i) <t(i+1). Next, we solve the problem with dynamic programming using Bellman method (1), including a solution for unique optimum vector x*(i). We demonstrate that it will allow us to arrive at a singular solution to the problem in Appendix 3.

## 7. DFA as Universal NFT Auction

The current version of the DFA is designed to address the issue of the fractional (collective) ownership of an asset offered for the auction sale. DFA algorithms select the winning combination of bids to provide the maximum benefit to the seller. At the same time, the DFA provides additional functionality, making it possible with a minimum design effort to accomplish the variety of tasks addressed by other auctions designs that do not provide fractional ownerships capabilities.

- DFA can also be used for multi-unit auctions.
- DFA can have a single winner. The auction is reduced to a regular closed first-price auction if the parameter K is set to 1.

- In addition, if the seller also wants to participate in the ownership of the item, they can regulate the number of owners by changing the value of K and include it in the contract defining terms of collective ownership. This is another important difference from multi-item auction.

Specifically, DFA can also be used for the multi-unit auctions and can have a single winner. Moreover, DFA allows the participants to form groups before the auction begins and utilizes "all-or-nothing" type of order, to bid for all available tokens (K). If their bid is successful, it will distribute tokens between the group members as per group forming agreement. This capability is similar to what the PartyBid marketplace provides, which is making the PartyBid auction a particular case of the DFA's more general solution.

In an expanded version of DFA auction, participants can place restrictions on price and quantity of the fungible tokens while having the option of adding another condition in their bid to define a characteristic of a group they seek to form or belong. For example, if a participant is bidding for the partial ownership of a painting, they likely prefer to be in the winning group whose majority of members are other experienced collectors. Consider an example for the art collector category versus no category preference. The inputs are as follows:

n - number of bids.

i - bid index defining the order of bids entering the system from 1 to n.

$b_i$ - bid with the index i which is assigned when a bid is accepted by the system.

$c_i$ - category (class) selected for bid i (c=1 if the art collector class is selected, and c=0 for no category preference).

$w_i$ - number of tokens entered for bid i or bi, where wi is a whole positive number.

$p_i$ - price for w tokens entered for bid i or bi where p is a whole positive number.

$t_i$ - time stamp for bid i when it is accepted by the system.

K – the maximum number of Fungible Tokens that the NFT for sale has been fractionalized into.

$A = \{i \mid c_i = 1\}$,

$B = \{i \mid c_i = 0\}$.

$X_i$ - set to 1 if a $b_i$ is selected for the winning group and is set to 0 if the bi is not selected for winning group.

**Problem definition**:

Maximize the sum of the selected bids:

$\max \sum_{i=1}^{n} P_i * X_i$, satisfying the following conditions:

1. Total sum of tokens of the elected bids does not exceed the maximum number of fractionalized tokens K:

$$\sum_{i=1}^{n} W_i * X_i \leq K.$$

2. If total sum of $X_i$ belonging to set A equals

$$\sum_{i \in A} X_i > 0,$$

then it is required that the condition of majority is satisfied as well:

$$\sum_{i \in A} W_i * X_i > \sum_{i \in B} W_i * X_i.$$

3. To obtain a unique solution, if there are two or more aggregate bids that are equal in price, priority is given to the one that is the earliest, which is determined by comparing the times of the latest bids in the competing groups. If the latest bids are the same, then the comparison moves back to a bid before last, and so on (see the example in Appendix 1).

# Appendix 1. Selecting a Unique Group of Winners

Example 1: K=5, IBI=20

b=(w,p), where w is the weight, p is the price selected by the last bid in the group:

B={b1=(4,10), b2=(2,4), b3=(1,2), b4=(1,3), b5=(5,11), b6=(4,7), b7=(2,8), b8=(3,4), b9=(3,5), b10=(2,3), b11=(1,1),

b12=(2,6), b13=(1,2), b14=(5,10), b15=(1,4), b16=(1,2), b17=(3,3), b18=(1,3), b19=(3,5), b20=(5,11)}.

Two groups with equal maximum combined price:

Group 1= {b7=(2,8), b12=(2,6), b15=(1,4)} F* = 8+6+4=18 K*= 2+2+1=5. Group 2= {b7=(2,8), b4=(1,3), b15=(1,4), b18=(1,3)}    F* = 8+3+4+3=18  K*=2+1+1+1=5.

The winner is Group 1, selected on the basis of the last bids. b15 is an earlier bid than b18.

Example 2: K=6, IBI=21.

Modification of the previous example when the last bids in the competing groups are the same - add one new order b21= (1, 5) to the list.

Two groups with equal maximum combined price:

Group 1={b7={(2,8), b12=(2,6), b15=(1,4), b21=(1,5)}   F* = 8+6+4+5=23 K*=2+2+1+1=6.

Group 2 ={b4=(1,3), b7=(2,8), b15=(1,4), b18=(1,3), b21=(1,5)} F* = 8+3+4+3=23  K*= 2+1+1+1+1=6.

The winner is Group 1 selected on the basis of the bid before last. b15 is an earlier bid than b18.

## Appendix 2. Pruning Input

Step 1. Calculate the boundaries:

$$K(w) = \lfloor K/w \rfloor$$

Step 2. Consider class B(w) that contains all objects of weight w and order elements in B(w) by price $p_i$ in descending order. Choose from B(w) bids starting from the largest price until the amount of the gathered bids does not exceed K(w). In case, when prices are equal, choose bids with lesser t. This is done for each class B(w) ⊂ B. We shall solve a problem with bids only from these new classes a reduced problem.

Note: To satisfy the condition of proper ordering, a reduced set of bids should be reordered in accordance with the time of arrival t before applying Bellman algorithm.

## Appendix 3. Bellman Method

We solve the problem with dynamic programming using Bellman method, including a solution for vector x*(i).

Let F*(I, j) denote a maximum possible value for the seller, which is obtained as a solution for a problem with parameters n=i and K=j, which is the value function for the sub-problem with parameters n=i and K=j, and the respective parts of the input array. Then recursive Bellman equation presents as follows:

F*(i, j) = max (F*(i-1, j), F*(i-1, j-w(i) + p(i)).

Following the recursive formula with initial conditions F*(0, w) = 0 and F*(i, 0) = 0, we calculate all the elements including F*(n, K). The process involves dual cycles, first an external cycle with the step 1 from i=1 to i=n, and then an internal cycle of all the weights from j=0 to j=K. The process ends with obtaining the value function maximum F*(n, K). Now we need to determine the winning group of bids which is a vector x* containing the maximum sums of prices. We start this process with an element from the array F*(n, K*) which we denote F*(i', K') such that:

F*(i', K') = F*(n, K*), and F*(i'-1, K*) < F*(n, K*).

The element i' is the earliest element in its group of winning bids. Let us call it x*(i') =1, the first element of the winning group. The variable w, or the current weight, becomes w(i') and the element i' is the earliest element in the bid array that changes the F* value. This provides the lexicographic order versus time of the bid entering the system and, consequently, the uniqueness of the solution. Obtained values for K' and i' now become the starting point for the next step of the process. We find the sum of the weights of selected elements for each step, and the process ends when the sum of the current weights equals K*.